\documentclass[aps,prb,twocolumn,superscriptaddress]{revtex4-1}

\usepackage{hyperref} % hyperreferences
\usepackage{graphicx}
\usepackage{amsmath}
\usepackage{amssymb}

\DeclareGraphicsExtensions{.png,.jpg,.eps}
\usepackage{xcolor}

\hypersetup{pdfstartview={XYZ null null 1.3},
	pdfpagemode=UseNone
	}

\begin{document}

\author{J. L. Lado}
\affiliation{Department of Applied Physics, Aalto University, Espoo, Finland}
\affiliation{Institute for Theoretical Physics, ETH Zurich, 8093 Z{\" u}rich, Switzerland}

\author{Oded Zilberberg}

\affiliation{Institute for Theoretical Physics, ETH Zurich, 8093 Z{\" u}rich, Switzerland}

\title{Topological spin excitations in Harper-Heisenberg spin chains}

\begin{abstract}
Many-body spin systems represent a paradigmatic platform for the realization of emergent states of matter in a strongly interacting regime. Spin models are commonly studied in one-dimensional periodic chains, whose lattice constant is on the order of the interatomic distance.  However, in cold atomic setups or functionalized twisted van der Waals heterostructures, long-range modulations of the spin physics can be engineered.  Here we show that such superlattice modulations in a many-body spin Hamiltonian can give rise to observable topological boundary modes in the excitation spectrum of the spin chain. In the case of an XY spin-$1/2$ chain, these boundary modes stem from a mathematical correspondence with the chiral edge modes of a two-dimensional quantum Hall state. Our results show that the addition of many-body interactions does not close some of the topological gaps in the excitation spectrum, and the topological boundary modes visibly persist in the isotropic Heisenberg limit.  These observations carry through when the spin moment is increased and a large-spin limit of the phenomenon is established.  Our results show that such spin superlattices provide a promising route to observe many-body topological boundary effects in cold atomic setups and functionalized twisted van der Waals materials.  
\end{abstract}

\date{\today}

\maketitle

\section{Introduction}

%%%%%%%%%%%%%%%%%%%%%%%%%%%%%%%%%%%%%%%%%%%%%%%%%%%
\begin{figure}[t!]
\includegraphics[width=\columnwidth]{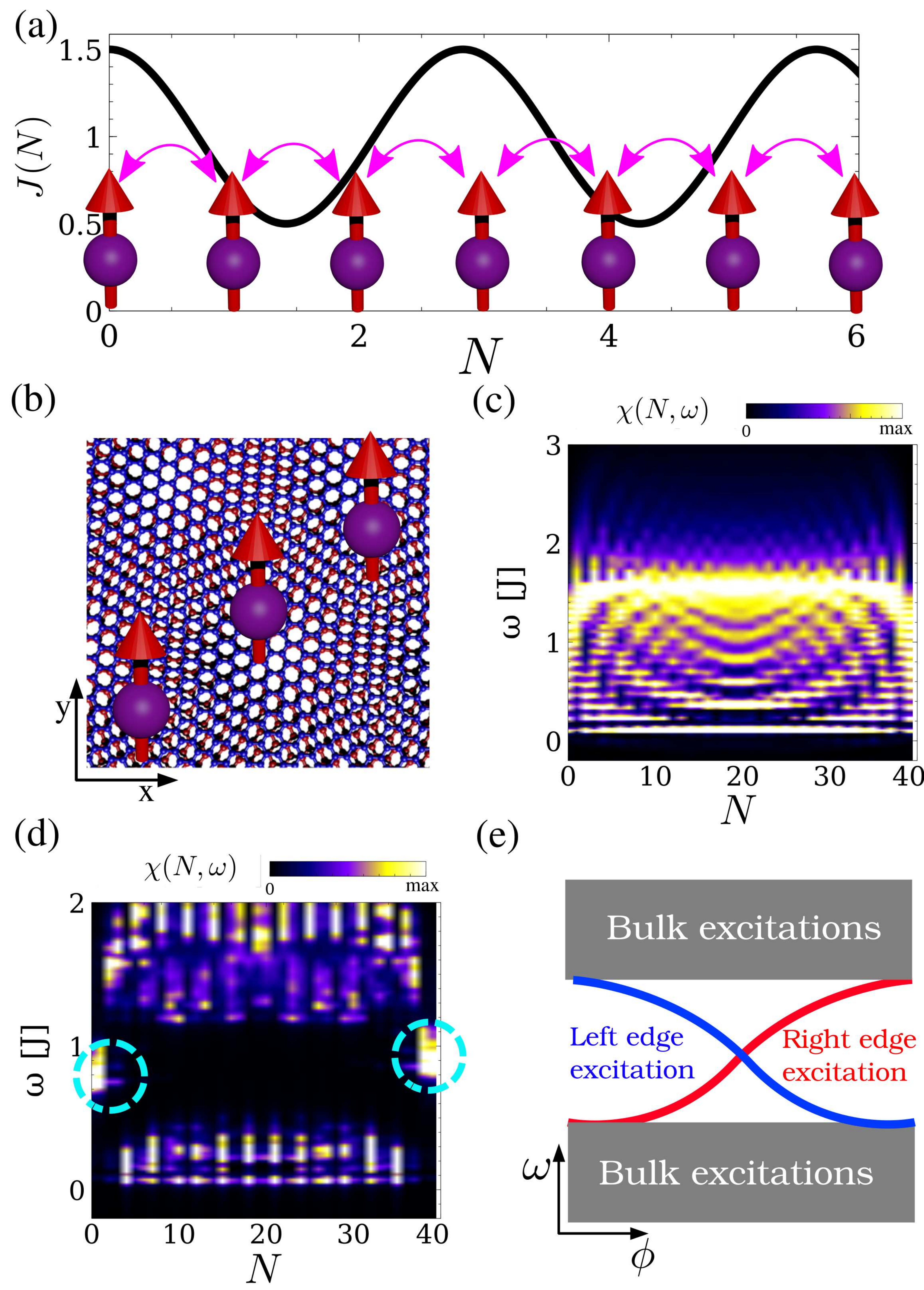}
\caption{
	(a) Sketch of a spin chain superlattice with Heisenberg couplings
	(magenta arrows)
modulated in space [cf.~Eq.~\eqref{modH}] that can be
	engineered in a  cold atom setup.
	Moreover, the model of panel (a) naturally arises
	in atomically engineered lattices
	on top of twisted van der Waals materials
	as shown in panel (b).
	(c) The calculated dynamical structure
factor [cf.~Eq.~\eqref{dynStruct}] for a uniform
	$S=1/2$ chain, showing the emergence of different
confined modes and a gapless excitation spectrum. (d) The calculated dynamical structure
factor of a modulated chain with $\alpha = \pi/\sqrt{2}$, $\lambda=0.5$
	and $\phi=0.6\pi$, showing the emergence of a spectral
gap and in-gap edge excitations (cyan circles). As the pumping parameter $\phi$
	is varied, we expect topological in-gap modes
to traverse the bulk gap as sketched in panel (e). 
}
\label{fig1} 
\end{figure}
%%%%%%%%%%%%%%%%%%%%%%%%%%%%%%%%%%%%%%%%%%%%%%%%%%%

Topological phases of matter comprise one of the most active research domains in contemporary physics research~\cite{RevModPhys.83.1057,RevModPhys.82.3045}. Prominent examples thereof involve
systems with translational symmetry, where characteristic topological boundary
effects appear\cite{RevModPhys.83.1057,RevModPhys.82.3045}, requiring in
certain scenarios an additional
symmetry constraint, such as chiral\cite{PhysRevLett.42.1698},
time reversal\cite{PhysRevLett.61.1329,PhysRevLett.95.226801,Qi2008}, or crystalline\cite{PhysRevLett.106.106802,RevModPhys.88.035005} symmetries.
Identification of this fundamental source for topology has enabled the realization of topological effects in a plethora of systems
including electronic~\cite{RevModPhys.83.1057,RevModPhys.82.3045}, 
photonic~\cite{RevModPhys.91.015006},
atomic~\cite{,RevModPhys.91.015005}, 
phononic\cite{Lu2009,SerraGarcia2018,PhysRevB.97.020102,Ssstrunk2015}
and circuit metamaterials.\cite{PhysRevLett.114.173902,PhysRevX.5.021031,Imhof2018,PhysRevA.82.043811}
Additionally, topological phenomena may arise
from single-particle interference in structures 
where competition between different length scales occurs, i.e., in structures
with broken lattice translation-invariance.
Examples of this include
breaking of translation-invariance with magnetic-fields in
integer quantum Hall effects~\cite{PhysRevLett.49.405}, 
topological pumps~\cite{PhysRevB.27.6083,Kraus2012a, Lohse2016, Lohse2018, Zilberberg2018,PhysRevLett.108.220401}, and topological quasicrystals~\cite{Kraus2012a, Kraus2012, Kraus2013, PhysRevB.91.064201,Kraus2016}. These systems share a deep connection with one another: an adiabatic time-dependent 
modulation between superlattice potentials sharing the same long-range order can
lead to topological pumping and a dynamical realization of
quantum Hall systems~\cite{PhysRevB.27.6083,Kraus2012a}.
%Interestingly, such topological phenomena are not intrinsic to the quantum world, and its implementations can exploit interference in purely classical systems.

The fundamental motivation for exploring the topology of spectral gaps in
physical systems is threefold: (i) nontrivial topology implies topological
phase transitions between systems of different topology, (ii) quantized bulk
responses appear in association with the topology, and (iii) at open boundaries,
the quantized topological bulk invariants lead to corresponding boundary
effects~\cite{RevModPhys.83.1057,RevModPhys.82.3045, RevModPhys.91.015006}. 
For example, the energy levels of the quantum Hall
effect are associated with topological invariants -- Chern numbers -- that
lead to the quantization of the bulk Hall conductance. In turn, with
open boundary conditions, the
quantum Hall effect exhibits corresponding chiral edge modes~\cite{RevModPhys.83.1057,RevModPhys.82.3045, RevModPhys.91.015006}. In similitude, topological pumps exhibit a quantized number of charges pumped per
pump-cycle corresponding to the Chern number of the pump~~\cite{PhysRevB.27.6083,Kraus2012a}. At their boundaries,
0D boundary modes must appear and cross the spectral gap during the per
pump-cycle.~\cite{Kraus2012a}

Moving towards the design of topological phenomena in many-body
quantum systems,\cite{Senthil2015} 
we consider a variety of platforms including cold
atom in optical superlattices\cite{Aidelsburger2014, Boada2015,
Mancini2015,Stuhl2015,Li2016,Goldman2016, PhysRevLett.122.110404}, and
atomically engineered lattices with scanning tunnel
microscopy~\cite{2019arXiv190409941C}. These
systems allow for the engineering of tailored quantum spin
models~\cite{Loth2012,Otte2008,Choi2017,PhysRevLett.114.076601,Toskovic2016,Spinelli2014,Bryant2015,PhysRevLett.119.227206,Choi2017,Baumann2015,Natterer2017}. A particularly versatile candidate in this direction
consists of hydrogenated graphene\cite{Yazyev2010}, where each hydrogen atom
binds an $S=1/2$ state in graphene\cite{Gonzalez-Herrero2016}. A key feature of
this system, relevant to our work, is that the system can be placed on top of
another graphene layer to form a moir\'e pattern that in turn leads to a long-ranged modulation of the spin-chain's
exchange-couplings\cite{Brihuega2017,PhysRevB.99.245118}. 
In particular, one can consider a single graphene layer where hydrogen
atoms 
are deposited equidistantly from each other.
By placing such a functionalized
graphene layer on top 
of another pristine graphene sheet and at
a relative angle, a moire pattern will
appear 
that effectively modulates the spin chain's exchange constants.
These
platforms offer the opportunity to study novel phenomena such as many-body
quantum phase transitions\cite{PhysRevLett.120.175702,
PhysRevLett.83.3908,PhysRevLett.86.1331,PhysRevB.65.014201,PhysRevB.71.134408,PhysRevB.66.052419,PhysRevLett.94.077201,PhysRevB.64.134419,Roux2013,
PhysRevA.78.023628,PhysRevB.89.161106,PhysRevB.98.104203,PhysRevB.91.104203} and many-body
localization\cite{PhysRevB.96.104205,PhysRevLett.115.180401,PhysRevB.87.134202}.

In this work, we show that topological boundary modes emerge in the
dynamical response spectrum of many-body spin
chains with a superlattice modulation.
In particular, we focus on
isotropic spin chains with modulated exchange
constants, that can be realized both in
cold-atom setups and the solid-state
platform discussed above. We harness a combination of a kernel polynomial
method~\cite{RevModPhys.78.275} with
tensor network techniques~\cite{itensor} to
compute the dynamical
structure factors of the
many-body system, that exhibit
the appearance of topological boundary excitations. We furthermore provide an analytic
adiabatic connection to known regimes, showing that in certain paradigmatic
cases, the topological modes can be adiabatically connected to a well-understood non-interacting
limit. Last, by a systematic scaling-up of the spins' moment, we obtain that our results persist also in the large-spin limit, showing the universality of such excitations in superlattices.

The manuscript is organized as follows: in Section~\ref{secfree},
we present the topological bulk and boundary effects corresponding to the
mapping between free particles
in one-dimensional superlattices, topological pumps, and two-dimensional
quantum Hall states. This section shows the connection between single
particle topological modes and many-body topological
response functions.
The next sections deal exclusively with topological
many-body response functions,
in systems
that cannot be mapped to a 
single particle picture.
In Sec.~\ref{secmany}, we detail the kernel polynomial
method~\cite{RevModPhys.78.275} and its utility in showing the emergence of topological
boundary excitations in many-body $S=1/2,1,3/2,2$ superlattice chains.
We, thus, reveal the existence of topological gaps in the excitation spectrum
of many-body superlattices. Finally, in Sec.~\ref{seccon}, we discuss and summarize our
results.

\section{Topological pumps and their boundary modes}
The main goal of this work is to show that 1D many-body superlattices
support topological gaps with in-gap 
boundary modes in their excitation spectrum. These
topological modes arise from the competition between different length scales in
the system. In conventional fermionic systems, such modes are
associated to a quantized topological pumping response in the bulk.
In particular, in the non-interacting limit, these boundary modes support the
quantized charge pumping in a finite
system~\cite{PhysRevB.27.6083, Kraus2012a}. Furthermore, they
correspond to the sampling over the chiral edge modes of a parent 2D quantum
Hall-like system~\cite{Kraus2012a}. Here,
we aim to extend such a mapping to strongly
interacting systems, by expressing the existence of topological boundary modes in a many-body framework.

We consider a Heisenberg model with a long-ranged modulation of its exchange
constants,\cite{PhysRevB.90.035150,Hu2015}
see Fig.~\ref{fig1}(a). 
Here we focus on spin models whose exchange constants are of the form
\begin{equation}
H=J \sum_N [1+\lambda\cos(\alpha N+\phi)]\vec{S}_N\cdot\vec{S}_{N+1}\,,
\label{modH}
\end{equation}
where $\vec{S}_{i}$ are spin operators with spatially-modulated coupling of
amplitude $\lambda$, modulation frequency $\alpha$, and displacement $\phi$.
The site index $N$ goes from $N=0$ (the leftmost site)
to $N=L-1$ (the rightmost site).
We note that the impact of non-periodicity on the
ground state of quantum Heisenberg models
has been addressed in the past.\cite{PhysRevLett.90.177205}
As stated previously, the Hamiltonian Eq.~\eqref{modH} can be
realized in cold atomic setups and solid-state platforms based on atomically
engineered
twisted 2D materials [Fig.~\ref{fig1}(b)].\cite{PhysRevLett.90.177205,PhysRevLett.93.037205,PhysRevB.99.245118,Brihuega2017,2019arXiv190602711L}
In the solid state realization of this Hamiltonian
based on hydrogenated twisted bilayer graphene,
the parameter $\alpha$ in
will be controlled by
the ratio between the hydrogen-hydrogen distance and the moire
length, whereas the parameter $\phi$ will be controlled by
the displacement between
the two layers.
When $\lambda=0$, the model describes a uniform antiferromagnetic Heisenberg
chain. Taking spins $S=1/2$, the spin chain is known
to have a gapless excitation spectrum, and represents a
paradigmatic integrable system that can be solved using Bethe's antsaz~\cite{betheansatz}.
For arbitrary $\lambda$ and $\alpha$, however, the system has no known solution.

In the following, we will show that a superlattice Hamiltonian of the form Eq.~\eqref{modH}
hosts topological boundary modes in its excitation spectrum. The existence of these boundary modes can be seen in the dynamical structure factor
\begin{equation}
\label{dynStruct}
	\chi(N,\omega)=
	\langle GS|S_N^z\delta(\omega-H+E_{0})S_N^z|GS\rangle\,,
\end{equation}
where $S^z_N$ is the spin operator along $z$ 
at the site number $n$, $E_0$ is the
many-body ground-state energy, and
$|GS\rangle$ is the many-body ground state of the system. 
We note that analogous dynamical structure factors
can be defined by taking operators for the different
spin components, so that the previous one correponds to
the $zz$ dynamical structure factor $\chi \equiv \chi^{zz}$.
This quantity is sensitive to
the spectrum of excitations in the system that are accessible by a local
perturbation at position $N$.
The details of the method to compute Eq. \ref{dynStruct} are detailed
in Sec. \ref{kpmdmrg}.
Using the dynamical structure factor, we can for example readily verify
the aforementioned gapless excitation spectrum property of Eq.~\eqref{modH} 
for the $S=1/2$ uniform antiferromagnetic Heisenberg
chain, see Fig.~\ref{fig1}(c). \footnote{Dynamical responses in the different
sites of the chain
that are not averaged over $\phi$ are denoted with the color scheme
of Fig. \ref{fig1}(c)}
Taking $\lambda\neq 0$, gaps appear in the
dynamical structure factor spectrum and  topological boundary modes can be
observed in the excitation spectrum of the boundary, see Fig.~\ref{fig1}(d).
In particular, the in-gap modes wind through the bulk excitation gap as a
function of $\phi$, see Fig.~\ref{fig1}(e).
The origin of such in-gap modes can be understood by starting from
a modulated non-interacting limit, as we show in the next section.

\subsection{Single-particle topological pumps and their bulk-boundary
excitation spectrum}
\label{secfree}
\begin{figure}[t!]
\includegraphics[width=\columnwidth]{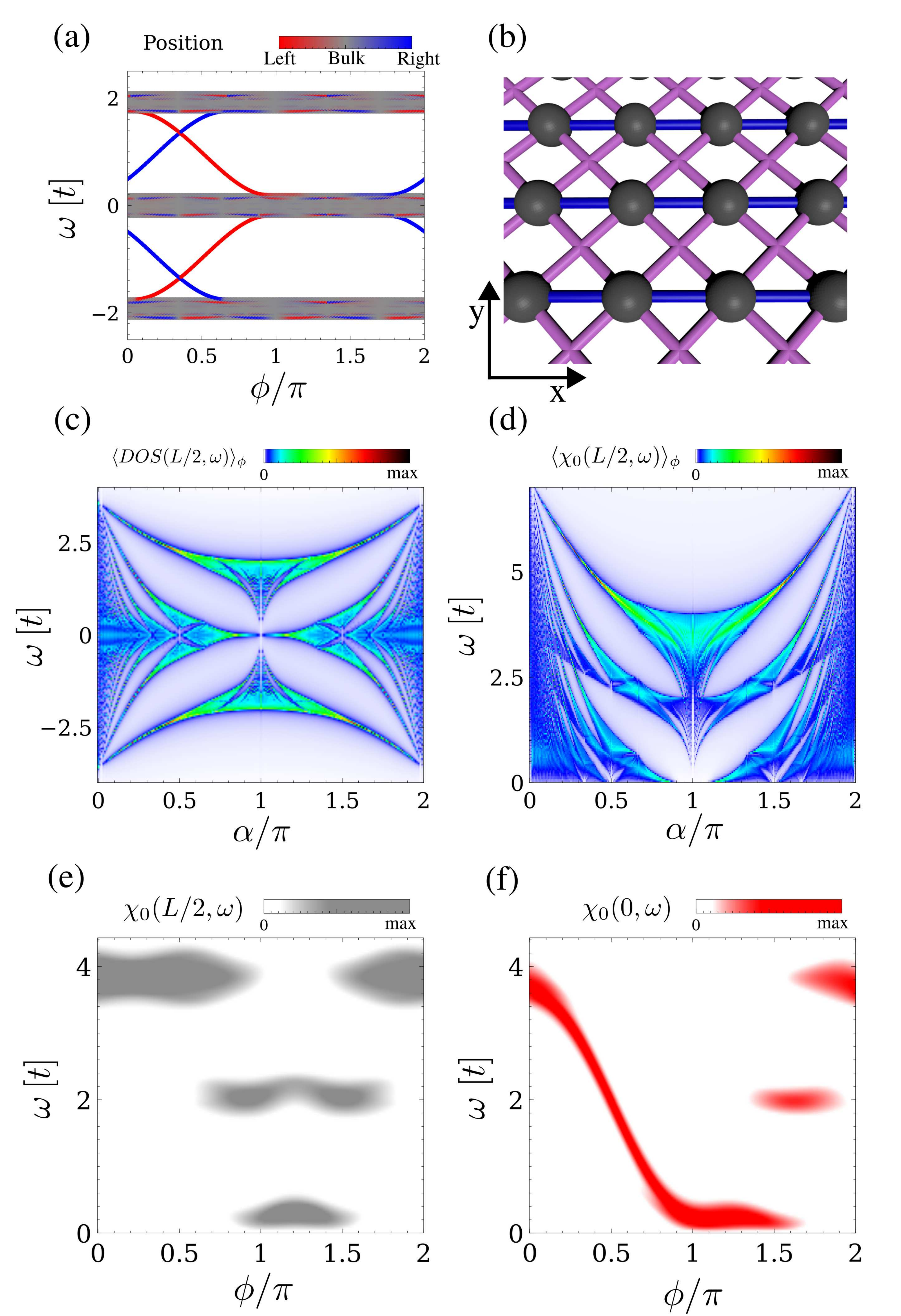}
\caption{
	Single-particle superlattices. 
	(a) The energy spectrum of the off-diagonal Harper model [Eq.~\eqref{freeh}] as a function of the
	pumping parameter $\phi$. Topological in-gap
	bound states appear on both sides of the chain (red and blue), and
	traverse the gap. The appearance of these states is in correspondence
	with the topological bulk response of the topological pump and can be
	mapped to the chiral edge modes of the 2D quantum Hall effect~\cite{Kraus2012a}.
	(b) Sketch of the 2D quantum Hall model [Eq.~\eqref{ed:HodH_2D}] that
	is mappable to the off-diagonal Harper pump [Eq.~\eqref{freeh}]. 
The Hofstadter spectra in a bulk site averaged over $\phi$ (c) and 
	the corresponding dynamical charge response in a bulk site averaged
	over $\phi$ (d)
	[cf.~Eq.~\eqref{SPCorr}].
	Panels (c,d) highlight their characteristic topological spectral
	gaps for arbitrary $\alpha$.
	Note the different energy axis in (c) and (d)
	due to the fact that the dynamical correlator includes contributions
	from transitions that can have absolute energy $0\leq \omega \leq W$,
	where $W=\omega_{\rm max}-\omega_{\rm min}$ is the full bandwidth of
	the bulk spectrum (c).
	(e) There are bulk gaps that remain open in the dynamical charge
	response as a function of $\phi$. (f) Same as (e) but evaluated at the
	boundary, showing in-gap boundary excitations.
 We used $\lambda=0.8$ and $\alpha=\pi/\sqrt{2}$
	in (acdef),
	panels (c) and (d) are averaged over $\phi$.}
\label{fig2} 
\end{figure}

Before focusing on the many-body study
of excitations of spin chains [Eq.~\eqref{modH}],
we first consider a specific non-interacting limit of Eq.~\eqref{modH}. 
The strongly interacting Hamiltonian Eq.~\eqref{modH} can be modified by
breaking its rotational symmetry to obtain a spin chain with anisotropic
exchange
	\begin{align}
H(\Delta)=J \sum_N 
		[1+\lambda\cos(\alpha N+\phi)][{S}^x_N{S}^x_{N+1} 
		\nonumber \\
		+ {S}^y_N{S}^y_{N+1} + \Delta {S}^z_N{S}^z_{N+1}]\,.
\label{hamilxxz}
\end{align}
In the limit $\Delta=0$, Eq.~\eqref{hamilxxz} becomes the Harper-XY model~\cite{Giamarchi2003}
 $H(\Delta=0)=\sum_N[1+\lambda\cos(\alpha N+\phi)][{S}^x_N{S}^x_{N+1} + {S}^y_N{S}^y_{N+1}]
$, which can be analytically solved by means of Jordan-Wigner's transformation~\cite{Giamarchi2003}
 $S^-_N = e^{\sum_{i<N} c_i^\dagger c_i} c_N$ and $S^+_N = e^{\sum_{i<n}
 c_i^\dagger c_i} c^\dagger_N$, with $S^\pm_N = S^x_N \pm i S^y_N$.
 Specifically, using the transformation, the Hamiltonian becomes
\begin{equation}
	H = t\sum_N [1+\lambda\cos(\alpha N + \phi)]
	c_N^\dagger c_{N+1} + h.c.\,,
	\label{freeh}
\end{equation}
with $t = J/2$. The model Eq.~\eqref{freeh} in known as the off-diagonal Harper model,\cite{Harper1955,Kraus2012} 
which was used in the realization of photonic topological
pumps.\cite{Kraus2012a, Zilberberg2018} Specifically, it exhibits bulk gaps and
topological boundary modes that thread through the gaps as a function of a scan
of the pump parameter $\phi$, see Fig.~\ref{fig2}(a). 
The appearance of these in-gap modes stems from the
topological quantized bulk response of the pump, which can be traced back to a
two-dimensional quantum Hall model on a lattice using dimensional
extension.~\cite{PhysRevLett.49.405, PhysRevLett.71.3697,
RevModPhys.91.015006, Kraus2012a, Kraus2012,2019arXiv190504549M}

For completeness, we detail the relationship between the 1D topological pump
and the 2D QHE. Let us start with a two-dimensional quantum Hall tight-binding model with
nearest-neighbor hopping in the $x$-direction and next-nearest-neighbor
hopping along the $\pm x\pm y$-direction [see Fig.~\ref{fig2}(b)] 
\begin{align} \label{ed:HodH_2D}
    H =& \sum_{N,M} \Big[ t\, c_{N,M}^\dag c_{N + 1,M}
                + \frac{\lambda}{2} \Big( e^{i \alpha N} c_{N,M}^\dag c_{N+1,M+1} \nonumber \\
    & \qquad + e^{-i \alpha N} c_{N,M}^\dag c_{N+1,M-1} \Big) + h.c. \Big]\,.
\end{align}
We have written the model in the Landau gauge and used Peierls' substitution\cite{Peierls1933} to describe the  magnetic flux piercing each plaquette of the model. In this gauge, the model does not depend on $y$ explicitly and it can be written 
in terms of momenta $k_y$ as good quantum numbers, leading to a summation over Eq.~\eqref{freeh} with $k_y\equiv \phi$. In other words, superlattice Hamiltonians can be understood to be specific $k$-cuts of a two-dimensional Hall state, where the magnetic flux $\alpha$ competing with the lattice translation in 2D is mapped onto the superlattice modulation frequency in 1D.

For rational values of $\alpha/(2\pi)$, periodic boundary conditions can be found in the $x$-direction with an additional
momentum $k$. Correspondingly, a Chern number\cite{RevModPhys.82.1959} can be computed for occupied bands of the model
 $\mathcal{C} = \frac{1}{2\pi}\int \Omega_{\alpha k} dk d\phi$,
where $\Omega_{\alpha k} = i[\sum_{j\in \mathcal{O}}
\partial_k  \langle \Psi_j | \partial_\phi \Psi_j \rangle
-   \partial_\phi \langle\Psi_j | \partial_k \Psi_j \rangle ]$,
$\mathcal{O}$ denotes the set of occupied states, and $\Psi_N$
are the eigenstates of the system, which for simplicity we assume
to be non-degenerate. In the case of a degenerate spectra, the Chern number
can be efficiently computed by means of
the Wilson loop technique.\cite{PhysRevB.83.035108}

%%%%%%%%%%%%%%%%%%%%%%%%%%%%%%%%%%%%%%%%%%%%%%%%%%%%%%%%%%%%
%%%%%%%%%%%%%%%%%%%%%%%%%%%%%%%%%%%%%%%%%%%%%%%%%%%%%%%%%%%%

The existence of a non-zero Chern number for the pump [Eq.~\eqref{freeh}]
implies that for open boundary conditions the system will develop a
quantized topological bulk response\cite{PhysRevB.27.6083,PhysRevLett.49.405, PhysRevB.91.064201, SchulzBaldes1999,KELLENDONK2002,PhysRevLett.45.494}
with associated boundary modes, see Fig.~\ref{fig2}(a). In particular, the
modes that traverse the gap as a function of $\phi$ appear in pairs
that are located at opposite boundaries of the modulated chain, where the
number of such pairs equals the Chern number of the gap. In this way, for a
specific chain at a particular $\phi$, a certain in-gap state can be located
inside the gap, whose origin can be traced back to the
non-trivial Chern number of the parent Hamiltonian~\cite{Kraus2012a}. 
For arbitrary $\alpha$ and $\lambda$, the bulk of the non-interacting superlattice chain [Eq.~\eqref{freeh}]
will have a fractal hierarchy
of gaps with different Chern numbers~\cite{PhysRevLett.49.405,PhysRevB.27.6083,PhysRevB.62.R10618}.
Plotting the bulk density of states as a function of $\alpha$ shows these
different gaps forming the so-called Hofstadter butterfly spectrum\cite{PhysRevB.14.2239,PhysRevLett.75.1348},
see Fig.~\ref{fig2}(c). Since these gaps have non-trivial Chern numbers, the boundary of the system will host in-gap modes. 

Such a topological bulk-boundary correspondence analysis cannot be easily extended to many-body systems, due to the fact that direct access to all many-body eigenstates is in general not possible. In order to make the connection with a many-body system, it is useful to demonstrate the non-interacting limit using an response that can be easily defined
for a many-body system: a dynamical correlation function.

\subsection{Topological pumps in a many-body framework}
In a many-body system, the existence of topological boundary modes
is defined by means of dynamical quantities instead
of by single-particle eigenvalues.
As an example, let us consider the non-interacting Hamiltonian Eq. \ref{freeh}:
the charge-charge correlation carries information on the single-particle
spectrum of the system, and is analogous to the $ZZ$ spin correlator [Eq.~\eqref{dynStruct}] of the original
$XY$ model [Eq.~\eqref{hamilxxz}]. Therefore understanding the non-interacting
limit provides a fruitful starting point to understand the many-body case.

The onsite charge-charge dynamical correlator can be expressed as
\begin{equation}
    \chi_0(N,\omega)= \langle GS |
    c^\dagger_N c_N \delta (\omega - H + E_{GS}) c^\dagger_N c_N
    | GS \rangle\,,
		\label{SPCorr}
\end{equation}
where $E_{GS}$ is the many-body ground state energy
and $|GS\rangle$ the
many-body ground state wavefunction. For free fermions, the previous
charge-charge correlator can be computed
from the single particle orbitals and eigenvalues
of Hamiltonian Eq. \ref{freeh} by means of Kubo's
formalism as $
\chi_0 (N,\omega) \sim 
\text{Im} \left ( \sum_{\mu,\nu}\frac{f_{\mu,\nu}}{E_\mu-E_\nu - \omega + i0^+}\right )
$, with $f_{\mu,\nu} = |\Psi_\mu(N)|^2|\Psi_\nu(N)|^2 
[n(E_\mu)-n(E_\nu)]$, 
$n(x)$ the Fermi-Dirac distribution at $T=0$ (i.e. a step function)
and $\Psi_\mu$ the single-particle 
eigenstates corresponding to single-particle eigenenergies $E_\mu$.
We note that the filling of the free fermion model has to be taken
at half filling i.e., with the chemical potential at $\omega=0$,
which
is the situation mathematically equivalent to the spectral function of the
XY model. Calculations away from
half filling in the fermionic model are analogous to an XY model with
external magnetic field. We emphasize that the many-body response function
of Eq. \ref{SPCorr} can be thus characterized from the single particle
eigenvalues of Eq. \ref{freeh}, making a connection between a many-body
response function and single particle energies. 

The spectral weight of $\chi_0$ can be understood as a weighted convolution
of the density of states of the system. 
The response with the highest energy is expected
at $\omega \approx 4t$, since it corresponds
to transitions between the deepest occupied state (located at
$\omega \approx -2t$) and the
highest unoccupied state (located at $\omega \approx +2t$).
For a system showing
different gaps in its spectra, $\chi_0$ will exhibit this structure
in a convoluted fashion. We can now compute the bulk $\chi_0$ at position $L/2$ where $L$ is the length of the chain
and average over different $\phi$.
We plot $\langle\chi_0(L/2,\omega) \rangle_\phi=
\frac{1}{2\pi}
\int_0^{2\pi} \chi_0(L/2,\omega,\phi) d \phi$
in Fig.~\ref{fig2}(d) for the off-diagonal Harper model Eq.~\eqref{freeh}.
First, we observe
that the fractal gap structure in the energy spectra in Fig.~\ref{fig2}(c)
\footnote{Spectral functions averaged over $\phi$
will be denoted with the same color scheme as Fig.~\ref{fig2}(c)} 
is
indeed manifesting in Fig.~\ref{fig2}(d). Importantly, when the bulk gaps remains open, the in-gap excitations located at the boundary of the system generate a signal in the local response.
The latter is clearly seen in Figs.~\ref{fig2}(e)-(f), where we observe that inside a finite
excitation spectral gap of the bulk Fig.~\ref{fig2}(e), there are in-gap boundary
excitations in Fig.~\ref{fig2}(f) that cross
the gap as a function of $\phi$. These in-gap excitations stem
from the convolution of the original single-particle topological
pump modes shown in Fig.~\ref{fig2}(a).
As a result, the non-trivial boundary phenomenon
of the 1D topological pump 
is
observable in the dynamical charge susceptibility.

The previous formulation of pumping modes in terms of a dynamical response
has the advantage that it can be also defined for a purely many-body
system, where single-particle energies are no longer defined.
In the next sections we will explore in system
that no longer have single particle excitations, and thus require
to compute the dynamical response function from the many-body ground state
explicitly.
In the following, we will use such formulation to show the emergence
of topological edge modes
in modulated spin Heisenberg models, a paradigmatic example of
a modulated quantum many-body Hamiltonian.

\section{Boundary excitations of topological pumps in many-body systems}
\label{secmany}
The mapping between a one-dimensional cosine-like long-wavelength modulated Hamiltonian and a
two-dimensional quantum Hall state is valid in the free electron case. However,
for strongly interacting superlattice Hamiltonians, such a mapping cannot be
readily done. In the following, our main goal
is to 
address whether boundary effects corresponding to topological pumps appear
also in a many-body dynamical response for generic
modulated quantum Heisenberg models.

Excitations in one dimensional $S=1/2$ models are usually studied by means of
Jordan-Wigner's transformation and
bosonization techniques~\cite{Giamarchi2003}. In this framework, low-energy excitations are understood by means
of Luttinger liquid excitations~\cite{Giamarchi2003}. This approximation, however, holds only for small energies,
which makes predictions concerning high-energy excitations difficult. To treat high-energy excitations,
matrix-product techniques are very well suited, as they allow to exactly solve
one-dimensional Hamiltonians without relying on a low-energy
approximation. In the following, we harness a
combination of a tensor-network formalism together with kernel polynomial
techniques to compute dynamical structure factors, and show that superlattice
many-body systems can host topological boundary modes in their excitation
spectrum. We now elaborate on the numerical procedure that allows us to compute
the dynamical properties of the superlattice Heisenberg Hamiltonian
Eq.~\eqref{modH}.

\subsection{Dynamical correlators with the DMRG-KPM method}
\label{kpmdmrg}
The kernel polynomial method\cite{RevModPhys.78.275} (KPM) allows for the 
computation of the function $\chi$ directly in frequency space, by performing
expansion in terms of Chebyshev polynomials.
For simplicity, we focus our discussion on
a Hamiltonian $\bar H$ whose ground state energy
is located at $E=0$ and whose
excited states are restricted to the interval $[0,1)$,
\footnote{
The MPS-KPM algorithm can be performed
with a Hamiltonian whose full spectrum is scaled and shifted
to fit the interval $(-1,1)$, which
for computational efficiently should
be done in with the spectral center
located at 0.
}
which can be generically obtained by shifting and rescaling the original Hamiltonian $H \rightarrow \bar H$. The
dynamical correlator $\chi$ for the original Hamiltonian $H$
can then be recovered by rescaling back the energies in the
dynamical correlator $\bar \chi$ of the scaled Hamiltonian $\bar H$. 

The dynamical correlator $\bar \chi$ for the Hamiltonian $\bar H$ takes the form
\begin{equation}
\bar \chi (\omega)=\langle GS|S_N^{z}\delta(\omega - \bar H)S_N^z | GS\rangle\,,
\end{equation}
where $|GS\rangle$ is the many-body ground state of the system. To
compute the dynamical correlator, we perform an expansion of the form
\begin{equation}
\bar \chi(\omega)=\frac{1}{\pi\sqrt{1-\omega^{2}}}\left(\mu_{0}+2\sum_{l=1}^{N_P}\mu_{l}T_{l}(\omega)\right)\,,\label{KPM}
\end{equation}
where $T_l$ are Chebyshev polynomials. The coefficients of the expansion $\mu_l$
can be then computed as
\begin{equation}
\mu_l = \int_{-1}^{1} \bar \chi(\omega) T_l (\omega) d \omega\,,
\end{equation}
which can be rewritten as
\begin{equation}
\mu_{l}=\langle GS|S_N^zT_{l}(\bar H)S_N^{z}|GS\rangle\,.
\end{equation}

Taking into account the recursion relation of the Chebyshev polynomials
\begin{equation}
T_l (\omega) = 2 \omega T_{l-1} (\omega) - T_{l-2} (\omega)\,,
\end{equation}
with $T_{1}(\omega) = \omega$ and $T_0 (\omega) = 1$,
the different coefficients $\mu_l$ can be computed
by iteratively defining the vectors
\begin{eqnarray}
|{w_{0}}\rangle & = & S_N^{z}|GS\rangle\\
|w_{1}\rangle & = & \bar {H}|w_{0} \rangle\\
|w_{l+1}\rangle & = & 2\bar {H}|w_{l}\rangle-|w_{l-1}\rangle\,,
\end{eqnarray}
so that $|w_{l}\rangle=T_{l}(\bar H)S_N^{z}|GS\rangle$. 

In this way, the coefficients
$\mu_l$ are computed as 
\begin{equation}
\mu_{l} = \langle GS |S_N^z | w_l \rangle\,.
\label{KPMproj}
\end{equation}
To improve the convergence rate of the expansion, the coefficients are redefined $\mu_l \rightarrow g_l^{N_P} \mu_l$, using the 
Jackson Kernel\cite{Jackson1912}
$g_{l}^{N_P}=\frac{(N_P-l-1)\cos\frac{\pi l}{N_P+1}+\sin\frac{\pi l}{N_P+1}\cot\frac{\pi}{N_P+1}}{N_P+1},$
to damp Gibbs oscillations~\cite{RevModPhys.78.275}. 
The number of polynomials used $N_P$ controls the natural
smearing of the $\delta(x)$ function,
yielding a smearing that scales as $1/N_P$ in units of the whole bandwidth.
In particular, the bigger the number of polynomials $N_P$, the sharper
the spectral features will be.
Given that the bandwidth of the full Hamiltonian scales as $S^2L$, the
smearing in units of the exchange coupling scales as $S^2L/N_P$.
As a reference, we took up to $N_P=4000$ for the $S=1/2$ calculations,
and $N_P=60000$ for $S=2$ calculations.
With these coefficients,
the dynamical structure factors in the whole frequency range can be computed
with the same resolution
using Eq.~\eqref{KPM}. The previous procedure can be used also to compute
dynamical correlators between different sites simply by replacing the operator
$S_N^z$ in Eq.~\eqref{KPMproj}. 

Importantly, the KPM-workflow can be readily implemented
within the matrix product state
formalism\cite{PhysRevLett.69.2863,RevModPhys.77.259,PhysRevB.87.155137} 
using ITensor~\cite{itensor}, that enables us\cite{dmrgpy} to compute the
dynamical correlation function of many-body systems
directly in frequency space.\cite{PhysRevB.90.115124,PhysRevB.91.115144,PhysRevB.92.115130,PhysRevB.90.045144} In the following, we demonstrate the power of this method in identifying boundary modes of topological pumps in the excitation spectrum of different spin superlattices.

\subsection{Boundary excitations of topological pumps in S=1/2 chains}
\label{sec:s12}

%%%%%%%%%%%%%%%%%%%%%%%%%%%%%%%%%%%%%%%%%%%%
\begin{figure}[t!]
\centering \includegraphics[width=1\columnwidth]{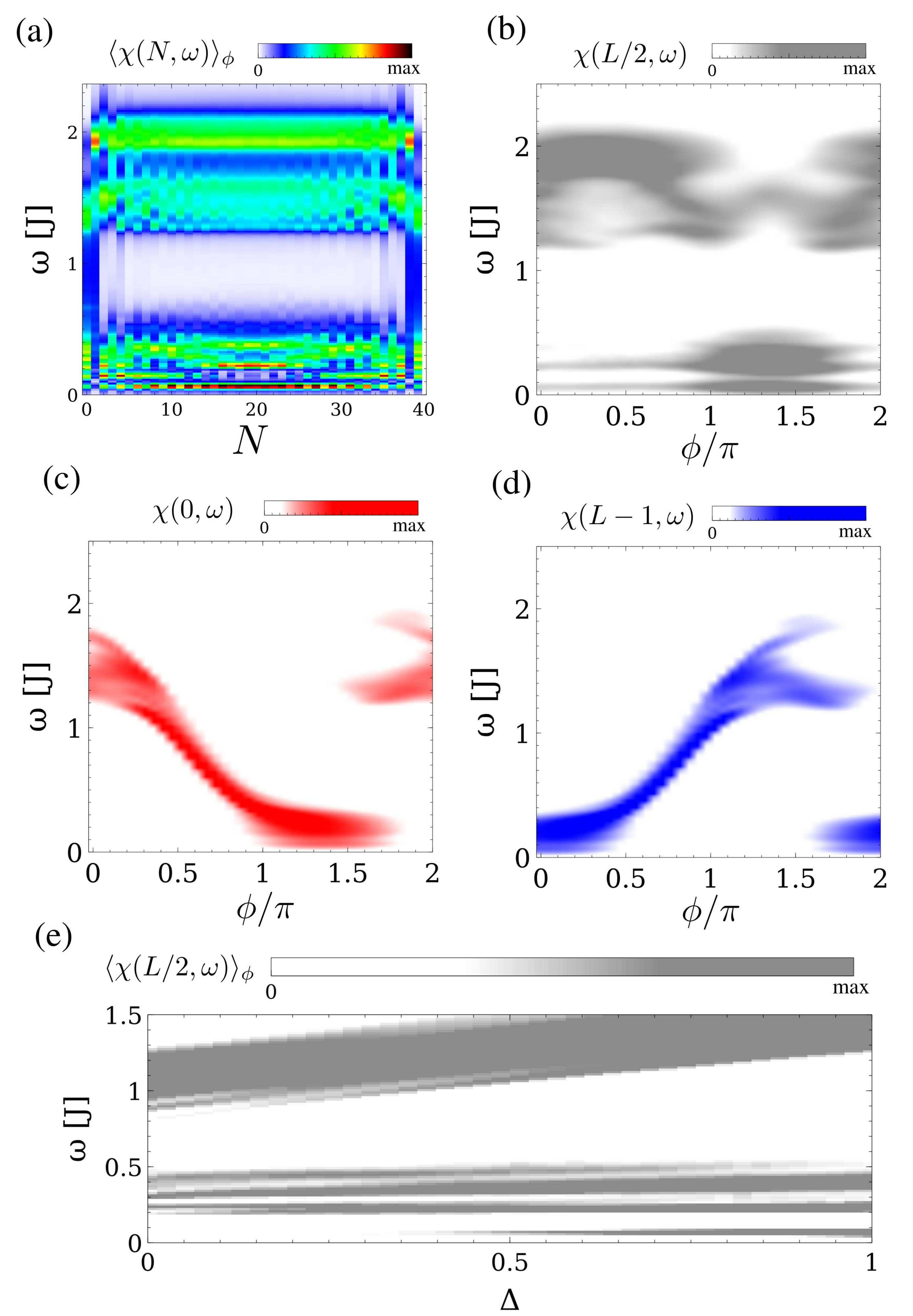}
	\caption{
		(a) Dynamical structure factor 
	$\chi(\omega)$
	[Eq.~\eqref{dynStruct}]
	in the different sites of the chain averaged over $\phi$
	for a Heisenberg spin chain $S=1/2$ of Eq.~\eqref{modH}.
	It is observed that the bulk of the chain shows a spectral gap,
	whereas at the edge the average spectral function
	$\langle \chi(\omega) \rangle_\phi$
	signals the appearance of edge excitations.
	This can be explicitly seen by
	looking at the dynamical structure factor
	$\chi(\omega)$ in the
        bulk (b), left edge (c) and right edge (d) as a function of $\phi$,
        showing states that thread through the gap at the edge while the bulk
	shows a robust spectral gap at the high energy part of the spectrum.
	The edge modes that pump with $\phi$ can be adiabatically
	connected to the ones found in the
	non-interacting limit Fig.~\ref{fig2}.
	Panel (e) shows the bulk dynamical correlator in
	the anisotropic case $\Delta\ne 1$, showing that the main
	excitation
	gap remains open as one goes from the non-interacting
	($\Delta=0$) to the fully interacting ($\Delta=1$) limit.
	We took 
	$\lambda=0.5$
	for panels (a,b,c,d,e),
	$\alpha=0.66\pi$ for (a,e)
	and  $\alpha=0.7\pi$
	for (b,c,d).
}
\label{fig3} 
\end{figure}
%%%%%%%%%%%%%%%%%%%%%%%%%%%%%%%%%%%%%%%%%%%%

%%%%%%%%%%%%%%%%%%%%%%%%%%%%%%%%%%%%%%%%%%%%
\begin{figure}[t!]
\centering \includegraphics[width=1\columnwidth]{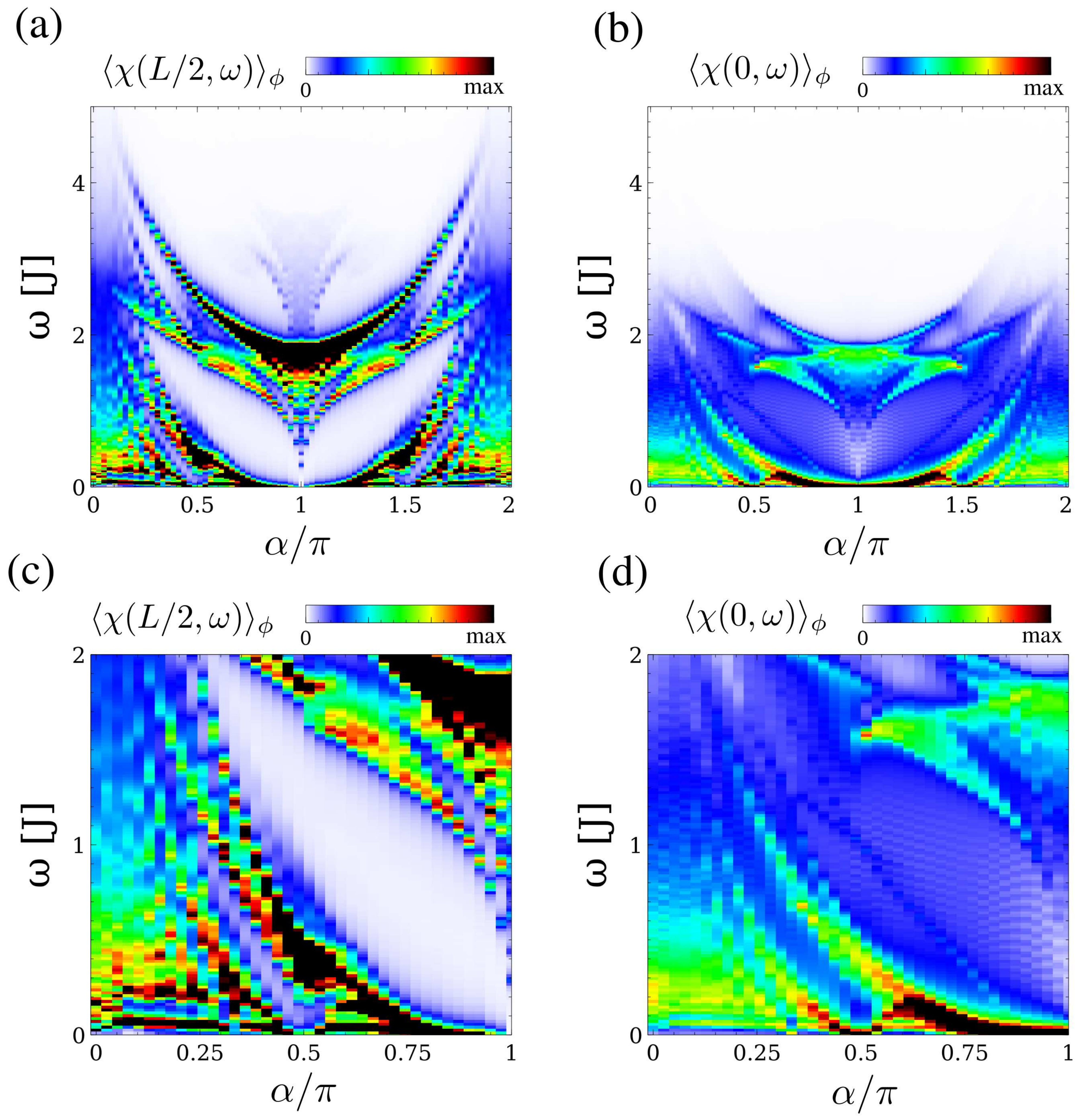}
\caption{
	(a-d) Dynamical structure factor 
	$\chi(\omega)$
        Eq.~\eqref{dynStruct}
	averaged over $\phi$,
	for a Heisenberg spin chain $S=1/2$ of 
	Eq.~\eqref{modH},
	for different modulation wavevectors $\alpha$.
	Panel (a) (zoom in (c)) shows the bulk and panel (b) (zoom in (d)) 
	shows the edge
	$\chi(\omega)$.
	It is observed that whereas the bulk (a,c) shows a spectral gap,
	the (b,d) shows a non-zero spectral weight, reflecting the
	emergence of the edge modes at arbitrary modulation frequencies
	$\alpha$.
	In the absence of $ZZ$ interaction in the
	Heisenberg model
	Eq.~\eqref{modH},
	panels (a,c)
	would be equivalent to panel Fig.~\ref{fig2}d.
	We took $\lambda=0.8$ for panels (a,b,c,d).
}
\label{fig4} 
\end{figure}
%%%%%%%%%%%%%%%%%%%%%%%%%%%%%%%%%%%%%%%%%%%%

We first focus on a spin chain with $S=1/2$, as it represents a minimal many-body system
whose topology can be adiabatically connected to a non-interacting limit discussed in Sec.~\ref{secfree}. We consider
a spin superlattice chain described by the Hamiltonian \ref{modH}.
Due to the rotational symmetry of Eq.~\eqref{modH}, the different
correlation functions are equivalent $\chi^{xx} (N,\omega)=\chi^{yy} (N,\omega)=\chi^{zz} (N,\omega)$,
which allows us to fully characterize the state by means of a single spin orientation
$\chi (N,\omega)\equiv \chi^{zz} (N,\omega)$.
Moreover, it is worth to recall the sum rule for the dynamical
correlator $\int \chi(\omega) d\omega = \langle GS | (S_N^z)^2 | GS \rangle$, which for
$S=1/2$ yields $\int \chi(\omega) d\omega=1/4$. This sum rule implies
that the spectral weight is conserved, and thus sites that yield a finite in-gap response must
compensate by decreasing their response in another energy window.

We compute the dynamical structure factor $\chi(N,\omega)$ [Eq.~\eqref{dynStruct}]
at each site of the chain, and average over different
pump parameters $\phi$, as shown in Fig.~\ref{fig3}(a). Whereas the bulk
of the system hosts an excitation gap, at the boundaries in-gap excitations appear. A more detailed picture
is obtained by comparing the dynamical structure factor in the bulk and at the boundary
as a function $\phi$, see Figs.~\ref{fig3}(b,c,d), respectively. In the bulk, we observe a gap in the
excitation spectrum [Fig.~\ref{fig3}(b)], whereas at the boundary
excitations cross the gap as a function of $\phi$ [Figs.~\ref{fig3}(c) and (d)].
This is the very same phenomenon that we detailed in the non-interacting case [Figs.~\ref{fig2}(e) and (f)],
showing that topological gaps of excited states of the non-interacting model
survive the onset of many-body interactions, and appear in the dynamical
response of the system.\cite{Mastropietro2018,PhysRevB.93.245154,PhysRevLett.112.176401}

The emergence of topological boundary modes as a function of $\phi$ happens
for arbitrary values of $\alpha$. First, let us recall that
in the non interacting limit, the value of $\alpha$ was associated
to the magnetic field of the parent two-dimensional Hamiltonian,
and thus edge modes appear for arbitrary values of $\alpha$.
By adiabatically connecting the non-interacting Hamiltonian
Eq.~\eqref{freeh} to Eq.~\eqref{modH} by means of 
Eq.~\eqref{hamilxxz}, we have observed
gaps that do not close that support in-gap boundary modes
in the fully interacting limit for
arbitrary values of $\alpha$. 
In particular, we show in
Fig. \ref{fig3}e
the bulk
spectral function
of Eq.~\eqref{hamilxxz} as a function of $\Delta$,
observing a bulk gap that remained open for
$\Delta \in (0,1)$.
In the strongly interacting limit of $\Delta=1$,
the existence of such bulk spectral gap and edge modes
can be observed
by computing the dynamical
structure factor as a function of $\alpha$
both in the bulk and at the edge as shown in Fig.~\ref{fig4}.
In particular, we observe that whereas the bulk shows a robust
spectral gap
[Figs.~\ref{fig4}(a) and (c)],
the boundary shows a finite spectral
density in that very same energy window
[Figs.~\ref{fig4}(b) and (d)],
highlighting the emergence of in-gap boundary modes
for arbitrary
modulation frequency $\alpha$.

The adiabatic connection between
the Heisenberg model and free fermions is done by means
of Eq.~\eqref{hamilxxz}, which in particular breaks
rotational symmetry in the Heisenberg Hamiltonian.
Such rotational symmetry breaking
makes the different dynamical correlators
$\chi^{zz}$
and
$\chi^{xx}$
quantitatively different.
This motivates us to consider a mapping that retains the spin rotational
symmetry between the non-interacting limit and the interacting one.
Interestingly, besides the
Jordan-Wigner mapping introduced in Sec. \ref{secfree}
the persistence of the topological boundary modes can be mapped
to completely different free model, namely a Harper-Hubbard model.
This additional mapping to a free fermionic system does
not break the rotational
symmetry of the Hamiltonian, in strike comparison
with Eq.~\eqref{hamilxxz}.

To perform the mapping to the Harper-Hubbard model, let us consider a fermionic model similar
to Eq.~\eqref{freeh}, but now for spinful fermions with an
onsite Hubbard interaction
\begin{equation}
	\centering
\begin{split}
	H_\phi = U \sum_N c_{N,\uparrow}^\dagger  c_{N,\uparrow}
	c_{N,\downarrow}^\dagger  c_{N,\downarrow} + \\
	t\sum_N [1+\lambda\cos(\alpha N + \phi)]
	c_{N,s}^\dagger c_{N+1,s} + h.c.
\end{split}
	\label{hubbard}
\end{equation}
For $U=0$, the imaginary part of the spin susceptibility
of the previous Hamiltonian can be written as
$
\chi_0(N,\omega,\phi) \sim
\sum_{\mu,\nu}\frac{f_{\mu,\nu}}{E_\mu-E_\nu - \omega + i0^+}
$, with $f_{\mu,\nu} = \langle \Psi_\mu | S^+_N | \Psi_\nu\rangle \langle
\Psi_\nu | S^-_N | \Psi_\mu\rangle 
[n(E_\mu)-n(E_\nu)]$ and $\Psi_\mu$ the different single particle
eigenstates of Eq. \ref{hubbard} for $U=0$. 
Note that the previous expression
is equivalent to the charge susceptibility in the absence
of symmetry breaking for the spinless chain presented in Sec. \ref{secfree}.
In particular, such susceptibility will have 
analogous properties as
the charge susceptibility of the spinless fermionic
chain in the non-interacting limit.
As a result, in the limit $U=0$ the spin response
can be understood in the same way as the spinless fermionic
free case.
For increasing values of $U$, the charge
fluctuations of the system develop a global gap
that scales with $U$, whereas the spin excitations are substantially
less affected due to spin-charge separation.
In particular, we observe that the high energy gaps in the
dynamical spin-spin correlator remain open up to large values of $U$.
In particular, for large values of $U$, we can perform a
Hubbard-Stratonovic transformation
to the Hamiltonian Eq.~\eqref{hubbard} and map it to the very same Hamiltonian
in Eq.~\eqref{modH}, with $J = 4 t^2/U$. As a result, the spin excitations
in the non-interacting limit adiabatically evolve towards the interacting limit,
and thus its topological properties can be once more
inferred from the non-interacting scenario.

In this section we have shown that the edge modes
of a modulated Heisenberg $S=1/2$ chain can be adiabatically
connected to the topological modes of a non interacting limit.
This connection can be made both by means of
a Jordan-Wigner mapping to an
interacting spinless fermion model,
or through a Hubbard-Stratonovic transformation
to a spinful Hubbard model.
Irrespective of the mapping, the
analytic connection highlights the topological
origin of the edge modes in the modulated
$S=1/2$ Heisenberg model. In the following
we address the next step in complexity,
namely a modulated $S=1$ Heisenberg model,
where the previous two mappings are
not trivially applied.

\subsection{Boundary excitations of topological pumps in S=1 chains}
\label{secs1}

%%%%%%%%%%%%%%%%%%%%%%%%%%%%%%%%%%%%%%%%%%%%%%%%%%%%
\begin{figure}[t!]
\centering \includegraphics[width=1\columnwidth]{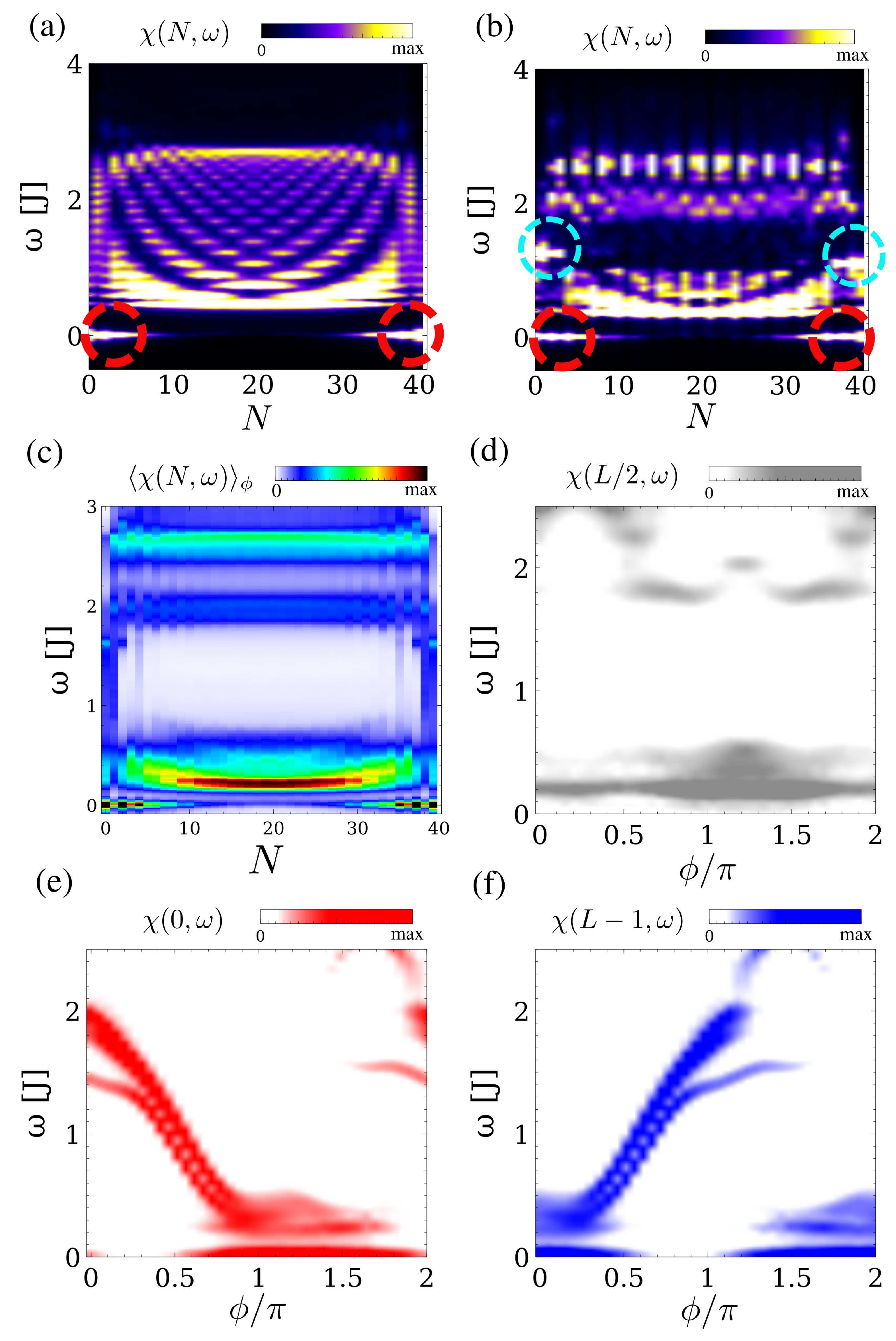}
\caption{
(a) Dynamical structure factor in the different
sites of the chain for a uniform $S=1$ Heisenberg model,
showing the emergence of topological modes at the edge
(red circles). Panel (b) shows the dynamical
structure factor for a modulated $S=1$ chain,
showing the coexistence of the preexisting zero modes (red circles)
with the topological pumping modes (cyan circles).
Panel (c) shows the dynamical structure factor
for the different sites of a $S=1$ chain averaged over $\phi$.
Bulk (d), left edge (e) and right edge (f) dynamical structure factors as a
	function of $\phi$,
showing states that thread through the gap in the edge
	while the bulk shows an excitation
	gap in the high energy part of the spectrum.
We took $\alpha=\pi/\sqrt{2}$ 
for panels (b,c,d,e,f),
$\lambda=0.4$ for panel (b) and $\lambda=0.7$ in 
panels (c,d,e,f).
}
\label{fig5} 
\end{figure}
%%%%%%%%%%%%%%%%%%%%%%%%%%%%%%%%%%%%%%%%

We now address the existence of boundary excitations associated with a topological pump
in a Heisenberg chain with $S=1$.
In striking comparison with the $S=1/2$ chains studied above, $S=1$ chains are much harder
to theoretically study as they cannot be easily connected to a non-interacting limit and, as a result,
we directly address the system using a full many-body formulation of
topological boundary modes. In the following, we show that
despite the missing non-interacting limit, modulated $S=1$ chains
show similar topological in-gap excitations.

It is instructive first to address the known
limit of $\lambda=0$, that corresponds to a uniform $S=1$ Heisenberg model. This
model is known to develop a bulk gap, which
has been shown numerically to converge to
a value of $0.41J$ in the thermodynamic
limit.\cite{PhysRevLett.69.2863} Moreover, such model develops
gapless edge modes,\cite{PhysRevLett.69.2863} namely, the Heisenberg model with
$S=1$ has the particularity
of hosting in-gap boundary modes that originate from its topological
non-trivial ground-state. Importantly, these modes  appear without the
requirement of a superlattice modulation, see Fig.~\ref{fig5}(a). As a result,
for weak superlattice modulations, the Hamiltonian can host simultaneously
boundary modes originating from the
original non-trivial topology of the uniform limit [red circles in Fig.~\ref{fig5}(b)], and also pumping boundary modes arising from the longer-ranged superlattice modulation [cyan circles
in Fig.~\ref{fig5}(b)].
For strong modulations, the original topological
gap of the uniform system closes and only the topological pumping modes of the superlattice survive.

We now proceed in an analogous way to the free electron limit [Sec.~\ref{secfree}]
and to the $S=1/2$ [Sec.~\ref{sec:s12}]. First, in Fig.~\ref{fig5}(c), we show the local dynamical correlator
at every site of an $S=1$ chain with open boundary conditions.
When averaged out over
the different phases $\phi$, 
the bulk of the $S=1$ spin chain shows an excitation gap as shown in Fig.~\ref{fig5}(c), alongside a finite spectral weight on the boundaries in that very same gap. This phenomenon
is the same as the
one observed for the $S=1/2$ chain [cf.~Fig.~\ref{fig3}(a)]. The
nature of the edge weight can be understood by looking at the $\phi$-dependent
dynamical structure factor. In particular, in the bulk, it is observed that a
spectral
gap appears for every $\phi$, see Fig.~\ref{fig5}(d). In comparison, at the boundary [Figs.~\ref{fig5}(e) and (f)],
we see a pumping in-gap excitation that traverses the gap as $\phi$ is varied. 

%%%%%%%%%%%%%%%%%%%%%%%%%%%%%%%%%%%%%%%%%%%%%%%%%%
\begin{figure}[t!]
\centering \includegraphics[width=1\columnwidth]{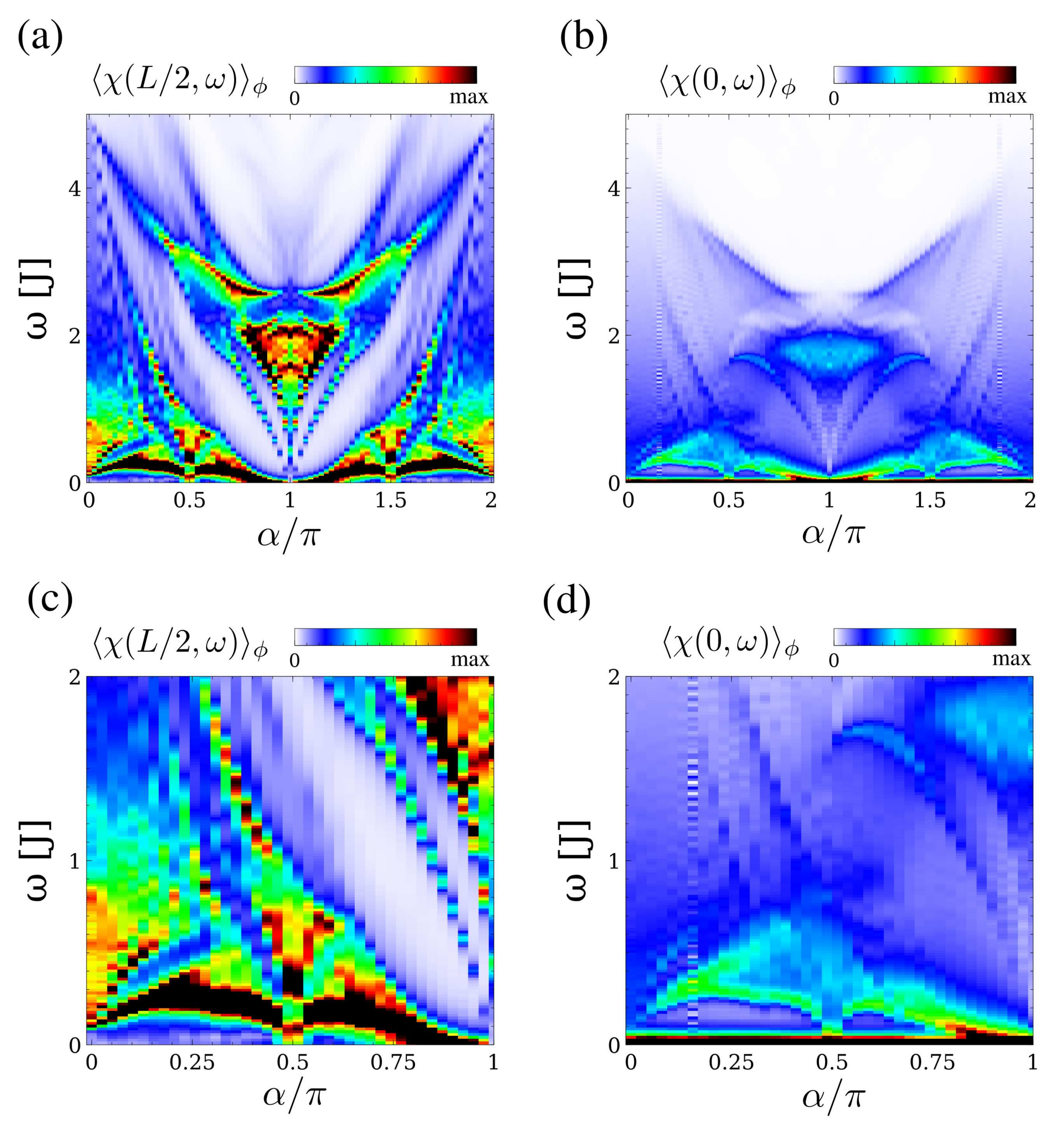}

\caption{
	(a-d) Dynamical structure factor
        $\chi(\omega)$
        Eq.~\eqref{dynStruct}
        averaged over $\phi$,
        for a Heisenberg spin chain $S=1$ of
        Eq.~\eqref{modH},
        for different modulation wavevectors $\alpha$.
        Panel (a) (zoom in (c)) shows the bulk and panel (b) (zoom in (d))
        shows the edge
        $\chi(\omega)$.
        It is observed that whereas the bulk (a,c) shows a spectral gap,
        the boundary (b,d) shows a non-zero spectral weight, reflecting the
        emergence of the boundary modes at arbitrary modulation frequencies
        $\alpha$.
	In contract with the $S=1/2$ chain of Fig.~\ref{fig4},
	the present case cannot be adiabatically connected
	to the free fermion Hamiltonian Eq.~\eqref{freeh}.
	We took $\lambda=0.8$ for panels (a,b,c,d).
}
\label{fig6} 
\end{figure}
%%%%%%%%%%%%%%%%%%%%%%%%%%%%%%%%%%%%%%%%%%%%%%%%%%

The existence of a high energy bulk
excitation gap together with in-gap edge modes emerges
for generic modulation frequencies of the Heisenberg superlattice.
This can be easily observed by computing the Hofstadter spectra for the modulated $S=1$ chain for different frequencies $\alpha$, see Fig.~\ref{fig6}. In particular, we see that a spectral gap appears
for a wide range of modulation frequencies $\alpha$ [Figs.~\ref{fig6}(a) and (c)]. For any of those frequencies, computing the structure factor at the boundary shows the existence of in-gap modes, see Figs.~\ref{fig6}(b) and (d).

For the $S=1$ studied above, no mapping to a free interacting limit can be easily performed.
Nevertheless, we identify topological boundary modes that traverse the gap in a similar fashion to that understood in the free fermionic topological pump limit. Given that in the strong interacting limit,
the computation of the Chern number cannot be performed, at this stage, it is not possible
to uniquely determine the invariant that protects these gap crossings. 

\subsection{Boundary excitations of topological pumps in high-spin chains}

%%%%%%%%%%%%%%%%%%%%%%%%%%%%%%%%%%%%%%%
\begin{figure}[t!]
\centering \includegraphics[width=1\columnwidth]{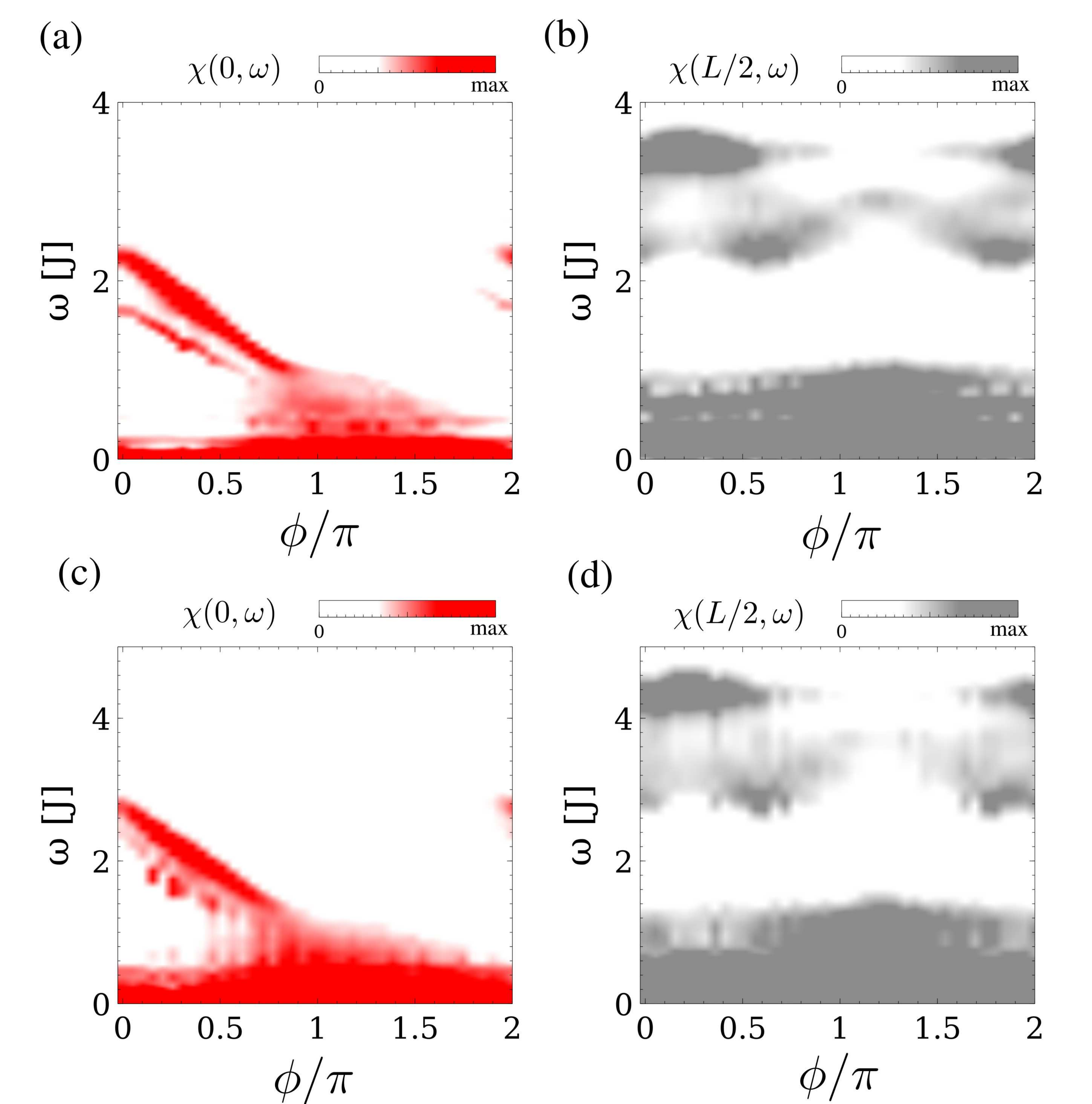}

	\caption{Dynamical structure factor $\chi (\omega)$ 
	at the edge (a,c) and bulk (b,d)
for a Harper-Heisenberg chain of
	Eq.~\eqref{modH}, for spins $S=3/2$ (a,b) and $S=2$ (c,d).
	In particular, panels
	(b,d) show the emergence of a spectral gap
	in the bulk, that hosts pumping
	modes at the edge (a,c).
	This situation is analogous to the pumping
	shown for $S=1/2$ in Fig.~\ref{fig3},
	$S=1$ in Fig.~\ref{fig5} and ultimately,
	the free fermion case of Fig.~\ref{fig2}.
	In comparison with the $S=1/2$
	case of Fig.~\ref{fig3}, a simple mapping
	with the free fermionic case of
	Fig.~\ref{fig2} can not be performed.
	We took $\alpha=\pi/\sqrt{2}$ and
	$\lambda=0.5$ for panels (a,b,c,d).
}
\label{fig7} 
\end{figure}
%%%%%%%%%%%%%%%%%%%%%%%%%%%%%%%%%%%%%%%

Previously, we focused on $S=1/2$ and $S=1$ chains,
for which we showed the emergence of topological pumping modes.
We turn to study whether such physics survives for higher-spin superlattice
chains,
and optimally, whether a large-$S$ limit
could be identified.\cite{2018arXiv181002626G}
Towards answering this question, we now study the case
of topological pumps for the $S=3/2$ and $S=2$ Heisenberg superlattice models,
following an analogous procedure as the one highlighted in the previous section.

We first point out several features
of the large-$S$ limit: the physics of large-$S$ spin chains resembles
in certain aspects the semiclassical limit. This can be qualitatively understood from the
fact that the commutation relation of the $S_\alpha$ matrices become less relevant
as the value of $S$ increases. In this regard, one could naively think that large-$S$
Heisenberg models would approach a classical limit with symmetry breaking, hosting
a N{\'e}el order. This is, however, not the case, as large-$S$ Heisenberg chains still retain
a singlet ground state with no symmetry breaking, and thus their ground state
must be treated within a many-body framework.\cite{PhysRevB.81.224404} In particular,
according to Haldane's conjecture
for integer $S$, the ground state of a uniform Heisenberg model is expected
to host a finite gap whereas for half-integer it is expected to be gapless,
which has been verified for $S=1/2$,\cite{Alcaraz1988,PhysRevB.12.3908},
$S=1$,\cite{PhysRevB.50.3037,PhysRevLett.69.2863}
$S=3/2$\cite{PhysRevLett.59.140,PhysRevLett.76.4955} 
and $S=2$.\cite{PhysRevLett.77.1616,Nishiyama1995,PhysRevB.60.14529}

We consider the Heisenberg superlatttice chains
of higher spin, focusing on $S=3/2$ and $S=2$
cases. We show in Fig.~\ref{fig7} the bulk and edge
spectra as a function of the pumping parameter $\phi$ for an $S=3/2$ and an $S=2$ spin chain,
which again show the emergence of boundary pumping edge excitations in bulk spectral gaps.
In particular, we observe
that the spectra for $S=3/2$ and $S=2$ are qualitatively similar
apart from an overall bandwidth increase. The bandwidth increase can be understood
in terms of an increase in the spin stiffness arising from the higher spin of the
chain. The similarity in the spectra suggests that the system is reaching
a large-$S$ limit, implying that the topological pumping states are a generic feature
of modulated quantum spin chains.

\section{Conclusion}
\label{seccon}
We have shown that quantum spin superlattice 
chains
host topological excitations originating 
from the mapping of the superlattice to a topological pump. 
Specifically, we have shown that the emergence of such boundary
modes in $S=1/2$ chains can be understood
using a continuous deformation into a free-particle superlattice,
where the 1D topological pump and its boundary modes are equivalent to a scan
over the physics of the integer 2D quantum Hall state. The fact that we can
perform this deformation between the 1D interacting Heisenberg model
and the 1D free-fermion case demonstrates that at excitation gaps that do not
close, the boundary in-gap excitations share the same topological origin. 
Such a deformation is verified numerically, showing bulk spectral
gaps that do not close as one adiabatically goes from
the free fermion limit to the Heisenberg limit.
This
is a first strong indication that the geometrical lengthscale competition
leading to nontrivial 
topology in single-particle models, carries on to the many-body world.
Crucially, we have shown that the very same topological boundary excitations appear in higher-$S$ spin chains, suggesting
that the emergence of topological
boundary modes is a generic feature
of superlattice Hamiltonians, even when an adiabatic connection to a
free-particle model is not known.

Our findings have several important consequences:
(i) our results motivate possible further extensions of
topological characterization to superlattice many-body systems and their
excitation gaps; 
(ii) we show that 
long-ranged spatial modulations in many-body 1D systems 
provide a platform to study topological effects
and their interplay with other many-body effects,
such as critical exponent and many-body localization; (iii) our results
highlight that modulated Heisenberg systems 
provide a compelling framework to explore the interplay of
topological pumping excitations and quantum magnetism; and (iv) using
contemporary numerical methods, we can explore a whole new range of many-body
phenomena corresponding to excitations far above common low-energy treatments.

\section*{Acknowledgments}
We acknowledge financial support from the Swiss
National Science Foundation. 
We thank M. Sigrist, M. Fischer, R. Chitra, 
M. Ferguson, S. Sack, P. Weber and F. Natterer for
fruitful discussions.
J. L. L. acknowledges
financial support from the ETH Fellowship program.

\bibliographystyle{apsrev4-1}
\bibliography{main}
 
\end{document}